\documentclass[fleqn,usenatbib]{mnras}

\usepackage{newtxtext,newtxmath}

\usepackage[T1]{fontenc}
\usepackage{ae,aecompl}
\usepackage{url}

\usepackage{graphicx}	
\usepackage{amsmath}	
\usepackage{pifont}
\usepackage{float}
\usepackage{color}

\newcommand{\nup}{$\nu_{\rm peak}^S$}

\usepackage{lineno}
\usepackage{ulem}

\title[PKS 1424+240: another masquerading BL Lac object
]{PKS\,1424+240: yet another masquerading BL Lac object as a possible IceCube neutrino source}

\author[P. Padovani et al.]{P. Padovani$^{1,2}$\thanks{E-mail:
ppadovan@eso.org}, B. Boccardi$^{3}$, R. Falomo$^{4}$, P. Giommi$^{5,6,7}$\\
$^{1}$European Southern Observatory, Karl-Schwarzschild-Str. 
2, D-85748 Garching bei M\"unchen, Germany\\
$^{2}$Associated to INAF - Osservatorio di Astrofisica e Scienza dello Spazio, Via Piero 
Gobetti 93/3, I-40129 Bologna, Italy\\
$^{3}$Max-Planck-Institut f\"ur Radioastronomie, Auf dem H\"ugel 69, 53121 Bonn, Germany\\
$^{4}$INAF - Osservatorio Astronomico di Padova, vicolo dell'Osservatorio 5, I-35122, Padova, Italy\\
$^{5}$Associated to Agenzia Spaziale Italiana, ASI, via del Politecnico s.n.c., I-00133 Roma, Italy \\
$^{6}$Institute for Advanced Study, Technische Universit{\"a}t M{\"u}nchen,
Lichtenbergstrasse 2a, D-85748 Garching bei M\"unchen, Germany\\
$^{7}$Center for Astro, Particle and Planetary Physics, New York University Abu Dhabi, 
United Arab Emirates\\
}

\date{Accepted XX. Received YY; in original form ZZ}

\pubyear{2022}

\begin{document}
\label{firstpage}
\pagerange{\pageref{firstpage}--\pageref{lastpage}}
\maketitle

\begin{abstract}
We show that the blazar PKS\,1424+240, which has been recently associated by IceCube with a 
neutrino excess at the $3.3\,\sigma$ level together with three other sources, is similar 
to the first plausible non-stellar neutrino source, TXS\,0506+056, in being also a masquerading 
BL Lac object, i.e., intrinsically a flat-spectrum radio quasar with hidden broad lines and a standard 
accretion disk. We point out that these two sources share other properties, including
spectral energy distribution, high powers, parsec scale properties, and possibly radio morphology. 
We speculate that the relatively rare combination of proton-loaded jets, possibly typical of 
high-excitation sources, and efficient particle acceleration processes, related to their relatively 
high synchrotron peak frequencies, might favour neutrino production in these two sources. GB6\,J1542+6129, 
which has also recently appeared twice in a list of IceCube associations, seems also to belong to this 
rare blazar sub-class, which includes at most $\approx 20$ {\it Fermi-}4LAC blazars. 
\end{abstract}


\section{Introduction}\label{sec:Introduction}

Nine years ago the IceCube Neutrino Observatory\footnote{\url{http://icecube.wisc.edu}} 
detected the first high-energy astrophysical neutrinos of likely extragalactic origin
with energies up to $> 1$ PeV ($10^{15}$ eV) and since then has produced a steady list of 
events \citep[e.g.][and references therein]{Aartsen2020}. At present, however, only 
two astronomical objects have been associated with a significance larger than 
$\sim 3\,\sigma$ with these astrophysical neutrinos. Namely, the blazar TXS\,0506+056 
\citep{icfermi,iconly} at $z=0.3365$ and the local ($z = 0.004$) Seyfert 2 galaxy 
NGC\,1068.
\cite{Aartsen2020}, in fact, have reported on an excess of neutrinos at the $2.9\,\sigma$ 
level from the direction of NGC\,1068 
and a $3.3\,\sigma$ excess in the northern sky due to significant p-values 
in the directions of NGC 1068 and three blazars: TXS\,0506+056, PKS\,1424+240, and 
GB6\,J1542+6129. 

\cite{Padovani_2019} showed that TXS\,0506+056, despite appearances, is
not a blazar of the BL Lac type but instead a masquerading BL Lac object, namely
a flat-spectrum radio quasar (FSRQ\footnote{Based on optical spectroscopy 
blazars are classified into FSRQs and BL Lac objects, with 
the former displaying strong, broad, quasar-like emission lines and the spectra of the latter 
being often completely featureless and sometimes exhibiting weak 
absorption and emission lines \citep[e.g.][]{UP95}.})
whose emission lines are swamped by a very bright, Doppler-boosted jet, 
unlike ``real'' BL Lacs, which are instead {\it intrinsically}
weak-lined. This is extremely relevant for two reasons: (1) ``real'' BL Lacs and FSRQs
belong to two very different physical classes, i.e., objects {\it without} and {\it
  with} high-excitation emission lines in their optical spectra, referred
to as low-excitation (LEGs) and high-excitation galaxies (HEGs),
respectively \cite[e.g.][and references therein]{Padovani_2017}; (2) 
masquerading BL Lacs, being HEGs, benefit from several radiation fields 
external to the jet (i.e., the accretion disc, photons reprocessed in the 
broad-line region (BLR) or from the dusty torus), which, by providing more 
targets for the protons might enhance neutrino production as compared to LEGs. 

The aims of this Letter are to: (1) investigate if the second blazar in the \cite{Aartsen2020}'s 
list, PKS\,1424+240, also qualifies as a masquerading BL Lac object; (2) check if 
TXS\,0506+056 and PKS\,1424+240 (and GB6\,J1542+6129) share other relevant properties.  
We use a $\Lambda$CDM
cosmology with Hubble constant $H_0 = 70$ km s$^{-1}$ Mpc$^{-1}$, matter
density $\Omega_{\rm m,0} = 0.3$, and dark energy density
$\Omega_{\Lambda,0} = 0.7$. Spectral indices are defined by $S_{\nu}
\propto \nu^{-\alpha}$ where $S_{\nu}$ is the flux at frequency $\nu$.

\section{The source}\label{sec:source}

\subsection{Main astronomical data}\label{sec:astro_data}

PKS\,1424+240 is one of the most distant TeV-detected blazars, ranked
number seven in TeVCat\footnote{\url{http://tevcat.uchicago.edu/}} at the
time of writing, and ``a rare example of a luminous HBL\footnote{Blazars are sub-divided
on the basis of the rest-frame frequency of their low-energy (synchrotron) hump (\nup) into
low- (LBL/LSP: \nup~$<10^{14}$~Hz [$<$ 0.41 eV]), intermediate- (IBL/ISP:
$10^{14}$~Hz$ ~<$ \nup~$< 10^{15}$~Hz [0.41 -- 4.1 eV)], and high-energy
(HBL/HSP: \nup~$> 10^{15}$~Hz [$>$ 4.1 eV]) peaked sources respectively
\citep{padgio95,Abdo_2010}.}''
\citep{Cerruti_2017}. Its \nup~has been estimated to be around $10^{15} -
10^{16}$ Hz, depending on its state \citep{Abdo_2010,Archambault_2014}.

\begin{figure}
\vspace{-0.6cm}
\includegraphics[width=0.5\textwidth]{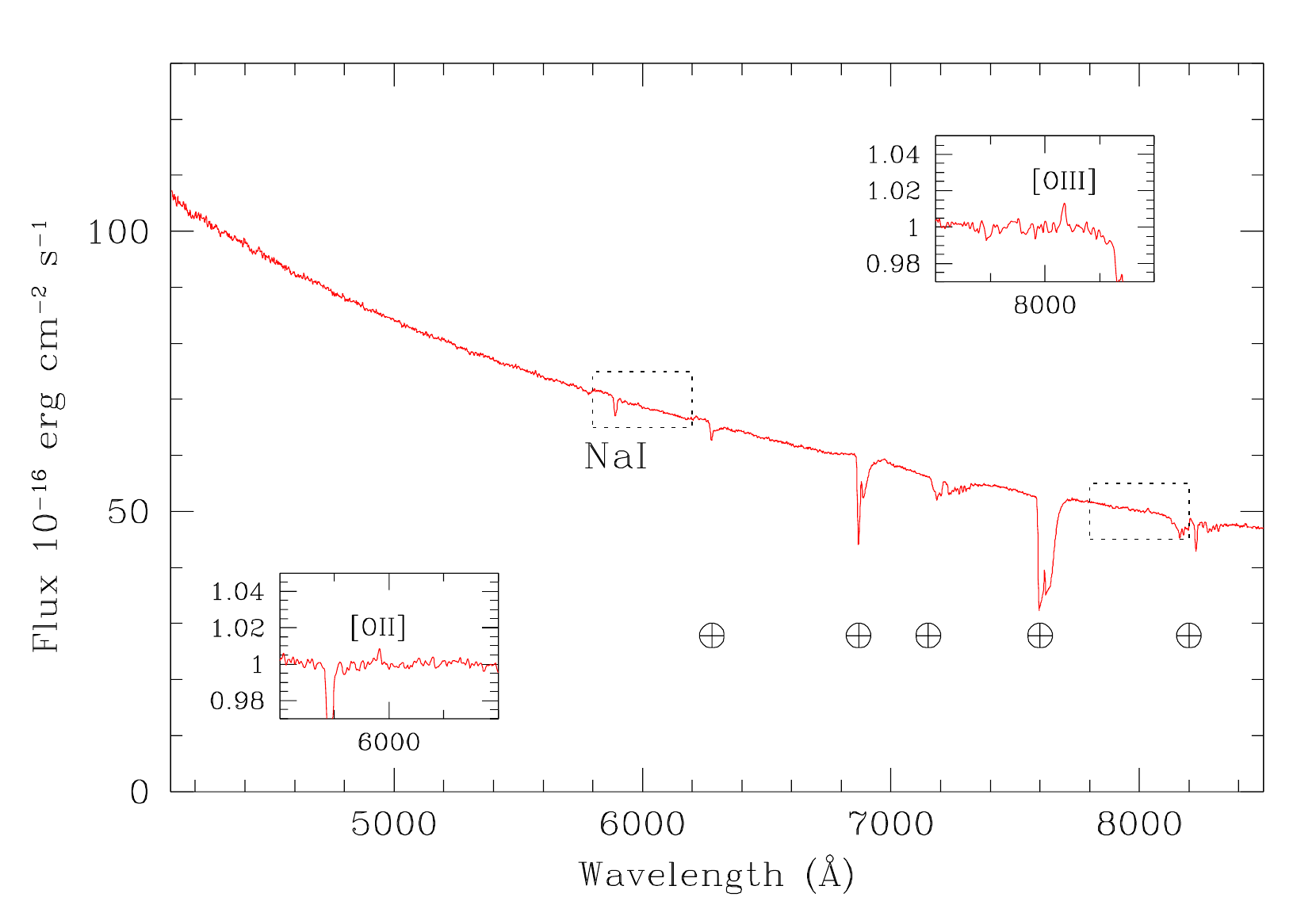}
\vspace{-0.7cm}
\caption{The optical spectrum of PKS\,1424+240 obtained at the Gran 
Telescopio Canarias with the OSIRIS spectrograph \protect\citep{Paiano_2017}. 
The continuum is well described by a power law with spectral index $\alpha = 
0.9$ (equivalent to F$_\lambda \propto \lambda^{-1.1}$). Two weak emission lines of 
[\ion{O}{II}] 3727 \AA~and [\ion{O}{III}]  5007 \AA~(see insets) are detected at z = 0.6047. 
Main telluric lines are indicated by $\oplus$. }
\label{fig:spectrum}
\end{figure}

The redshift of PKS\,1424+240 has a confusing history. The issue
has been settled by \cite{Paiano_2017}, who detected two faint emission
lines at 5981 and 8034 \AA~identified with [\ion{O}{II}] 3727 \AA~and
[\ion{O}{III}] 5007 \AA, with equivalent widths 0.05 and 0.10
\AA~respectively, at $z=0.6047$ (Fig. \ref{fig:spectrum}). This redshift
is also consistent with the presence of a group of galaxies at $z \sim 0.6$
very likely associated with the source \citep{Rovero_2016}. These 
detections imply line powers $L_{\rm [\ion{O}{II}]} \sim 4 \times 10^{41}$
erg s$^{-1}$ and $L_{\rm [\ion{O}{III}]} \sim 10^{42}$ erg s$^{-1}$.

As no sign of the host galaxy is apparent in the optical spectrum 
we cannot decompose it into 
a non-thermal power law and an elliptical galaxy template to estimate 
the black hole mass, $M_{\rm BH}$ \citep[e.g.][hereafter P22; see their Appendix for 
details]{Padovani_2022}. However, we
can place an upper limit on $L_{\rm \ion{Mg}{II}} < 5 \times 10^{42}$ 
erg s$^{-1}$ assuming this line has a full width at half maximum of 10,000 km s$^{-1}$.
By using eq. (5) and Table 2 of \cite{Shaw_2012} we then derive 
$M_{\rm BH} < 8 \times 10^{8}~\rm M_{\odot}$ (where $\rm M_{\odot}$ is one solar mass). 
Note that this is a 
very robust upper limit, only slightly above the value one would get 
assuming the host galaxy to be a typical giant elliptical 
($M_{\rm BH} \sim 6.3 \times 10^8 M_{\odot}$: P22),
which translates into an Eddington power $L_{\rm Edd} < 10^{47}$ erg 
s$^{-1}$, with $L_{\rm Edd} = 1.26 \times 10^{46}~(M/10^8 \rm 
M_{\odot})$ erg s$^{-1}$.


\subsection{Source characterization}\label{sec:source_char}

\begin{figure}
\vspace{-2.4cm}
\hspace{-0.6cm}
\includegraphics[width=0.5\textwidth]{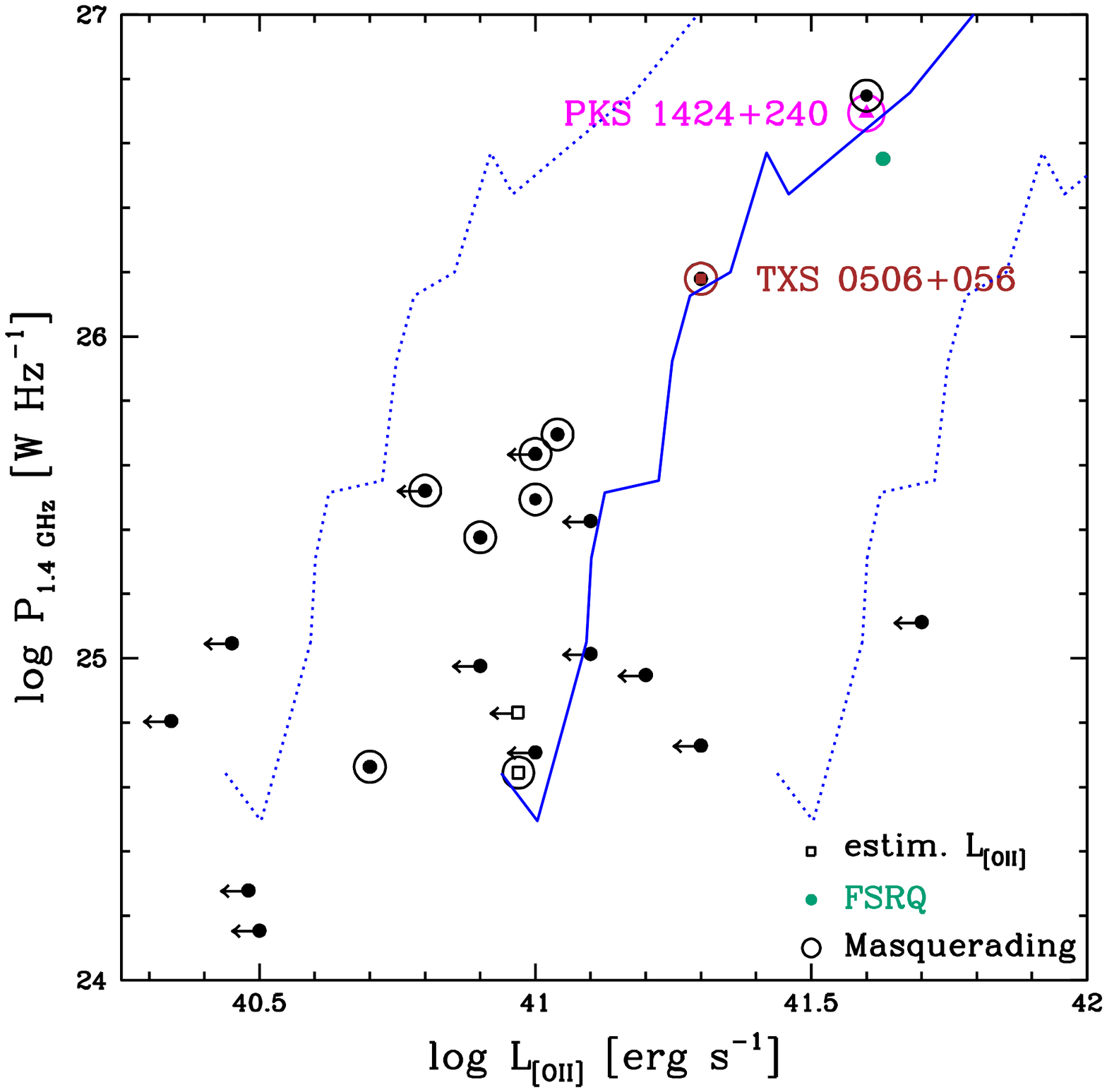}
\vspace{-3.0cm}
\caption{$P_{\rm 1.4GHz}$ vs. $L_{\rm [\ion{O}{II}]}$ for the P22's 
  sample (black
  filled circles), with masquerading sources highlighted (larger empty
  circles). Sources for which $L_{\rm [\ion{O}{II}]}$ has been estimated
  from $L_{\rm [\ion{O}{III}]}$ are denoted by black empty squares. The magenta 
  triangle is PKS\,1424+240, a brown circle indicates TXS\,0506+056, while the
  green filled circle denotes the single FSRQ in the sample. 
  The solid blue line is the locus of jetted (radio-loud) quasars, with
  the two dotted lines indicating a spread of 0.5 dex, which includes most
  of the points in Fig. 4 of \protect\cite{Kalfountzou_2012}.  
  Arrows denote upper limits on $L_{\rm [\ion{O}{II}]}$. Adapted
  from Fig. 1 of P22.}
\label{fig:Lr_LOII}
\end{figure}

We claim that PKS\,1424+240 is yet another example of a masquerading BL
Lac object linked to IceCube neutrinos. Following \cite{Padovani_2019} and P22 this
classification is based on four criteria:

\begin{enumerate}
    \item its location on the radio power -- emission line power, $P_{\rm
      1.4GHz}$ -- $L_{\rm [\ion{O}{II}]}$, diagram \citep[Fig. 4
      of][]{Kalfountzou_2012}, which defines the locus of jetted
      (radio-loud) quasars. With a catalogued 1.4 GHz flux density of 0.43
      Jy and a radio spectral index $\alpha_{\rm r} \sim 0.4$ (NASA/IPAC Extragalactic Database)  
      $P_{\rm
        1.4GHz} \sim 5 \times 10^{26}$ W Hz$^{-1}$, which, together with
      $L_{\rm [\ion{O}{II}]} \sim 4 \times 10^{41}$ erg s$^{-1}$, puts this
      source {\it exactly} on the locus, as shown by Fig. \ref{fig:Lr_LOII};
    \item a radio power $P_{\rm 1.4GHz} > 10^{26}$ W Hz$^{-1}$, typical of
      HEGs;
    \item an Eddington ratio $L/L_{\rm Edd} > 0.03$, hence $> 0.01$, which
      is also one of the defining characteristics of HEGs. This is the
      ratio between the accretion-related observed luminosity and the
      Eddington luminosity. The estimation of the former 
      is derived from $L_{\rm [\ion{O}{II}]}$ and $L_{\rm
        [\ion{O}{III}]}$, as detailed in \cite{Padovani_2019}. We obtain $L
      \sim 3 \times 10^{45}$ erg s$^{-1}$ and a BLR
      luminosity $L_{\rm BLR} \sim 10^{44}$ erg s$^{-1}$;
    \item a $\gamma$-ray Eddington ratio $L_{\gamma}/L_{\rm Edd} > 2$ and
      therefore above the proposed dividing line between ``real'' BL Lacs
      and FSRQs (0.1: \citealt{Sbarrato_2012}). In fact, $L_{\gamma} = 1.7 \times 10^{47}$ 
      erg s$^{-1}$ (0.1 -- 100 GeV), as derived from the 
      {\it Fermi}-LAT Fourth Source Catalog (4FGL-DR2) energy flux and photon 
      index \citep{4FGL,4FGL-DR2}. We stress that, due to its very large $L_{\gamma}$, 
      PKS\,1424+240 falls outside the boundaries of Fig. 2 of P22 
      (the $L_{\gamma} - P_{\rm 1.4GHz}$ plot) and is quite extreme in terms of its 
      location in Fig. 3 (the \nup~-- $L_{\gamma}$ plot) in the same paper. 
\end{enumerate}
In short, PKS\,1424+240 meets {\it all} the above criteria to be classified as a
masquerading BL Lac object.\\

\begin{figure}
\vspace{-0.4cm}
\includegraphics[width=0.48\textwidth]{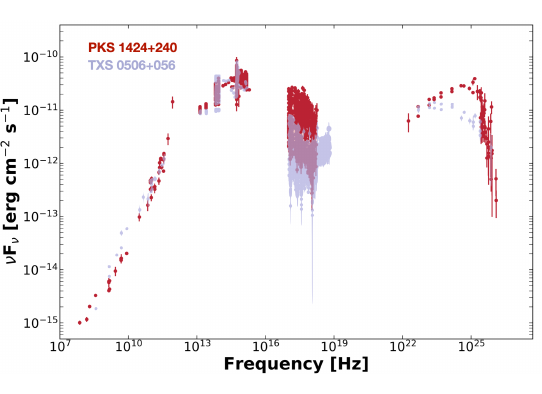}
\vspace{-0.8cm}
\caption{The archival SED of  PKS\,1424+240 (red points); 
for comparison the SED of TXS\,0506+056 is overlaid (light grey points). The overall SEDs of 
the two sources are quite similar. 
}
\label{fig:SED}
\end{figure}

\subsection{Similarities between PKS\,1424+240 and TXS\,0506+056}\label{sec:similarities}

Apart from being both masquerading BL Lacs, there are quite a few other commonalities between PKS\,1424+240 and TXS\,0506+056, which make them somewhat special and unusual blazars. Namely:
\begin{enumerate}
    \item they are both very powerful IBLs/HBLs, as TXS\,0506+056 has 
    $P_{\rm 1.4GHz} \sim 1.5 \times 10^{26}$ W Hz$^{-1}$ and $L_{\gamma} \sim 2.9 \times 10^{46}$ erg s$^{-1}$, $\sim 3 - 6$ times smaller than
    the PKS\,1424+240 values;
    \item the overall SEDs of the two sources are quite similar, both in normalisation 
    and shape. This is clear from Fig. \ref{fig:SED}, which shows the SED of PKS\,1424+240 
    (red) together with that of TXS\,0506+056 (grey). The low-energy parts of the SEDs are 
    almost identical, while there  are some difference in the X-ray and $\gamma$-ray bands, 
    mostly because of the higher average \nup~of PKS\,1424+240;
    \item their parsec scale properties, inferred from Very Long Baseline 
    Interferometry (VLBI) studies, are much more typical of HBLs than of  FSRQs. The observed brightness temperature of the 
    centimetre-VLBI core, $T_{\rm B}$, is in both cases relatively low, with median values 
    $\sim  6\times10^{10}$\,$\rm K$ in PKS\,1424+240 and $\sim 7\times10^{10}$\,$\rm K$ in 
    TXS\,0506+056 \citep{Homan_2021}, i.e. close to the equipartition 
    value (${\sim}5\times10^{10}$\,$\rm K$: \citealt{Readhead_1994}) and therefore indicative of a 
    modest Doppler boosting. This is also supported by proper motion measurements, revealing 
    low apparent speeds of $v_{\rm app}/c \equiv \beta_{\rm app} = 2.83\pm0.89$ \citep{Lister_2019} and $1.07 
    \pm 0.14$ \citep{Lister_2021} for PKS\,1424+240 and TXS\,0506+056 respectively. FSRQs, 
    especially the $\gamma$-ray powerful ones, are instead characterized by median core 
    $T_{\rm B}$ approaching and often exceeding the inverse Compton catastrophe 
    limit\footnote{The inverse Compton catastrophe is the rapid cooling through inverse Compton 
    scattering of a synchrotron-emitting region, which implies a threshold temperature 
    $\sim10^{12}$\,$\rm K$ \citep{Kellerman_1969}.}  due to strong Doppler boosting \citep{Kovalev_2009}, as well as by much higher maximum apparent speeds 
    ($\beta_{\rm app}\sim10 - 30$, e.g. \citealt{Jorstad_2017}).  \cite{Homan_2021}
    estimate relatively low Doppler and Lorentz factors, $\delta \sim 1.8$ and $\Gamma \sim 1.5$ 
    for TXS\,0506+056, and $\delta \sim 1.4$ and $\Gamma \sim 4$ and for PKS\,1424+240, as typically found for HBLs \citep[][and references therein]{Piner_2018}.\footnote{\cite{Homan_2021} derive  $\delta$ 
    by assuming that all sources have the same intrinsic median brightness temperature equal to the sample 
    median, $T_{\rm B,int} = 10^{10.609\pm0.067}$ K; hence $\delta = T_{\rm B,obs}/T_{\rm B,int}$. 
    This population-wide approach, while good for average values, might not be the most appropriate one for individual sources.}
A relatively low Lorentz factor is also consistent with the large apparent jet opening angles in both sources \citep[28$^{\circ}$ and 65$^{\circ}$ for TXS\,0506+056 and PKS\,1424+240, respectively, above the median value of 22.6$^{\circ}$ found for the $\gamma$-ray sources in the MOJAVE sample, see][]{Pushkarev_2017}. 
Indeed, the intrinsic aperture of the boosting cone is found to be inversely proportional to $\Gamma$, meaning that slower jets will appear less collimated. 
In conclusion, although these two sources have a standard (Shakura - Sunyaev) disk, as FSRQs, the VLBI properties of their jets are {\it not} FSRQ-like;

    \item their extended radio power, $P_{\rm ext}$, is relatively high, placing the sources in an intermediate regime between Fanaroff-Riley (FR) I and II objects \citep{Fanaroff_1974}, or rather closer to the FRII range ($\log P_{\rm ext}\gtrsim 25.5$ W Hz$^{-1}$ at 1.4 GHz). Very Large Array (VLA) imaging of PKS\,1424+240 at 1.4 GHz reveals a symmetric two-sided structure, with an extended flux density ${\sim}130\,\rm mJy$ \citep[Fig. 12,][]{Rector_2003}, which implies $\log P_{\rm ext}=26.3$ W Hz$^{-1}$ ($\alpha=0.8$), 
     at the lower end of FRII sources. No extended radio emission has been resolved in TXS\,0506+056 on kiloparsec (kpc) scales, but an estimate of the extended radio flux density can be obtained by comparing VLBI data with almost contemporaneous single-dish measurements obtained with the Owens Valley Radio Observatory telescope. \cite{Li_2020} have shown that the extended emission contributes between 1 and 18 per cent of the total, from which, since the smallest percentage
     is tied to the largest total flux density (and viceversa), we estimate an extended flux density in the range 15 -- 54 mJy i.e. $\approx 35$\,$\rm mJy$ at 15 GHz. This implies $\log P_{\rm ext}=25.9^{+0.2}_{-0.4}$ W Hz$^{-1}$ at 1.4 GHz for $\alpha=0.8$, which is close to the FRI/FRII break. Sources classified as BL\ Lacs but showing intermediate or FRII-like extended radio powers and morphologies are common among bright blazar samples \citep{Rector_2001,Kharb_2010}. However, this is mainly true for LBLs, while the properties of HBLs match those of their expected parent population, the FRI radio galaxies \citep{Rector_2001,Giroletti_2004}. For example, the HBLs in the Einstein Extended Medium Sensitivity Survey, characterised by $0.045 \leq z \leq 0.638$ ($\langle z \rangle \sim 0.3$), have $\log P_{\rm ext} \sim 23.7$ (and $\leq 25$) W Hz$^{-1}$ at 1.4 GHz \citep{Rector_2001}. 
     Therefore, PKS\,1424+240 and TXS\,0506+056 appear to be rare examples of IBLs/HBLs with FRII-like radio powers. If the extended radio power is a good proxy of the total jet power \cite[e.g.][]{Willott_1999}, the latter should also be in the FRII regime. High jet powers $\sim 10^{45} - 10^{46}$ erg s$^{-1}$ are also derived for both sources based on SED modeling (e.g. \citealt{Aleksic_2014, Ansoldi_2018}), much larger than the values, for example, of the low-redshift HBLs MKN 421 and MKN 501 ($\sim 6 \times 10^{43}$ erg s$^{-1}$: \citealt{Potter_2015});
   \item the issue of the radio morphology is, instead, difficult to address. TXS\,0506+056 is unresolved on kpc scales but the properties of the VLBI jet may be consistent with the development of an FRI morphology. In fact, the rapid increase of the jet opening angle with distance observed in VLBI images at 43 GHz has been interpreted by \cite{Ros_2020} as possible evidence for jet deceleration taking place at $\sim70-140$ pc from the BH, a process known to lead to the formation of an FRI jet \citep{Laing_2014}. The VLA image of PKS\,1424+240 shown by \cite{Rector_2003} presents mixed features, which would need to be better examined at higher resolution: the structure is reminiscent of an edge-brightened FRII, but the rough symmetry of the two-sided jet may be indicative of deceleration as observed in FRIs. In conclusion, one interesting possibility that needs to be further explored is the fact that PKS\,1424+240 and TXS\,0506+056 might belong to the rare class of efficiently accreting sources, which develop an FRI jet morphology, i.e., FRI-HEGs \citep[e.g.][]{Heckman_2014}. A sample of 1.4 GHz selected extended radio sources from the Sloan Digital Sky Survey with $0.03 < z < 0.1$ contains, for example, only 5/245, i.e. 2 per cent, of FRI-HEGs \citep{Miraghaei_2017}. Even if that were the case, however, these would have to be sources with FRI morphology and FRII-like jet powers, as discussed above. 

\end{enumerate}

\subsection{GB6\,J1542+6129}

\cite{Abbasi_2021} have performed a search for flare neutrino emission 
in the 10 years of IceCube data and reported a cumulative time-dependent neutrino excess 
in the northern hemisphere at the 3\,$\sigma$ level associated with only four sources: a 
radio galaxy, M87, two blazars, TXS\,0506+056 and GB6\,J1542+6129, and the Seyfert 2 galaxy 
NGC 1068, these last three sources being also associated with neutrinos by \cite{Aartsen2020}. 
GB6\,J1542+6129, an IBL \citep{Giommi_2021}, has then appeared twice in two years in a list of $3 - 3.3\,\sigma$ 
IceCube associations and is therefore also worth investigating. The redshift of this object is 
not available: all we know is that $0.34 \leq z \leq 1.76$ \citep{Shaw_2013}, 
which translates into $P_{\rm 1.4GHz} = 10^{26.4\pm0.7}$ W Hz$^{-1}$ and 
$L_{\gamma}/L_{\rm Edd} = 10^{0.3\pm0.9}$ (for $M_{\rm BH} = 6.3 \times 
10^8 M_{\odot}$, i.e. assuming the host galaxy to be a typical giant elliptical: Section \ref{sec:astro_data}). Both values are perfectly consistent with
a masquerading BL Lac classification (Section \ref{sec:source_char}). From the data given by 
\cite{Homan_2021} we derive $T_{\rm B}= 1.7 - 3.6 \times 10^{11}$ K, $\beta_{\rm app} = 
0.8 - 3.1$, $\delta = 4.2 - 8.7$, and $\Gamma = 2.3 - 5.0$ for the adopted redshift range. 
$\delta$ and $\Gamma$ are therefore also relatively small, which suggests that
GB6\,J1542+6129 shares similar radio properties with the other two sources.


\section{Discussion and Summary}\label{sec:discussion}
    The properties just described highlight the peculiarities of PKS\,1424+240 and TXS\,0506+056, which may be relevant in the context of neutrino emission models.
    In particular, a fundamental question concerns the jet particle composition, since the emission of neutrinos requires the presence of relativistic protons. The loading of heavy particles in the jet may occur either in the jet formation process or during jet propagation, as a consequence of entrainment of gas and stars from the environment. Jets formed through the Blandford-Znajeck mechanism \citep{Blandford_1977} are thought to be light, being mainly composed by electron-positron pairs, while disk-launched jets \citep[e.g.][]{Blandford_1982} are directly loaded with disk material, and are likely to be heavier and slower \citep[see also][]{Hawley_2006,McKinney_2006,Broderick_2015}. If both mechanisms take place at the same time in the AGN, mass exchange, possibly favoured by the development of magneto-hydrodynamic instabilities, may occur between the central relativistic jet and the surrounding heavy disk wind \citep[see e.g. the discussion by][]{Sikora_2007}. Mixing and entrainment of heavy particles from the ambient medium are also possible, particularly on larger scales when a jet decelerates and disrupts. Indeed, \cite{Croston_2018} were able to constrain the composition of the radio-lobe plasma in samples of FRI and FRII sources, showing that the decelerating jets in FRIs require a stronger proton content with respect to FRIIs. These authors also showed that this phenomenon is independent of the AGN accretion mode, being strictly related to the strong interplay between the jet and the environment, which characterizes FRI sources. 
    
    However, the most energetic phenomena in the jet, leading to the production of $\gamma$-rays up to the TeV regime, are known to occur at smaller distances from the BH, on sub-parsec and parsec scales \citep[e.g.][and references therein]{Madejski_2016}. Therefore, it is the jet composition closer to the central engine that may be most relevant in relation to neutrino production. Contrary to the scenario described above for the larger scales, modeling of the jet broadband emission \citep{Celotti_1993, Sikora_2005} and observational studies of the jet circular polarization \citep{Homan_2009} suggest that the dynamics of powerful jets is dominated by protons on VLBI scales. The parsec scale properties of the FRI jet in the LEG M\,87, on the other hand, are consistent with the dominance of a pair plasma \citep{Reynolds_1996, Broderick_2015} and, in general, the assumption of light jets appears adequate for the modeling of jet deceleration in FRIs \citep{Laing_2002, Laing_2014}. This possible difference in the jet composition may reflect a difference in the jet formation mechanism in LEGs (FRIs and FRIIs) and HEGs (mostly FRIIs and a few FRIs). 
    Based on the analysis of the jet collimation profiles, \cite{Boccardi_2021} have suggested that a more extended and prominent disk wind surrounds the relativistic spine in HEGs, while the jet expansion profiles in most LEGs are consistent with a jet origin in the vicinity of the ergosphere, as directly observed in M\,87. These results support the idea that jets in HEGs are more heavily loaded with protons on sub-parsec and parsec scales. 

It then follows that the production of neutrinos might be favoured in sources like PKS\,1424+240 and TXS\,0506+056. If jets produced by HEGs are indeed more heavily loaded with protons, and if protons and electrons are accelerated in the same regions and through the same mechanisms, then jets from IBL/HBL-HEGs may be those where the two conditions necessary for neutrino production are best met. Namely: 1) a high proton loading; 2) the occurrence of extremely efficient particle acceleration processes. In this scenario FSRQs would not fulfil the second requirement. In addition, as we mentioned above, the contribution from several external radiation fields typical of HEGs might further enhance neutrino production in these masquerading BL Lac objects.

Any physical mechanism, which may be invoked to explain the emission up to TeV energies in PKS\,1424+240 and TXS\,0506+056 will have to be reconciled, however, with the low $\Gamma$ and $\delta$ factors observed on parsec scales. Most solutions to the so-called {\it Doppler factor crisis} have been proposed so far for the case of ``classic'' low-power HBLs. These include the near-core decelerating jet models \citep[][and references therein]{Piner_2004} and the shock-in-jet models, in which the low apparent speeds are proposed to reflect the formation of quasi-stationary shocks in the jet \citep[][and references therein]{Hervet_2016}. It is a subject for future investigation whether such solutions are also applicable to the case of the high-power jets in PKS\,1424+240 and TXS\,0506+056.

In summary, PKS\,1424+240, as TXS\,0506+056, is another example of a masquerading BL Lac object associated with an IceCube 
neutrino excess. The two sources share also other properties, including spectral energy distribution, high 
powers, parsec scale properties, and possibly radio morphology. We suggest that the 
combination of proton-loaded jets, which might be typical of high-excitation sources, and efficient particle 
acceleration, implied by their relatively high synchrotron peak frequencies, might favour neutrino production. 
Finally, we note that
GB6\,J1542+6129, which has been also recently associated to neutrinos, seems also to belong to this very rare blazar 
sub-class, although a redshift determination would help to confirm this. 

Based on their very high radio and $\gamma$-ray powers we estimate that similar sources make up 
only 2.1 per cent of the total IBL and HBL population\footnote{This is the fraction of sources with radio and 
$\gamma$-ray powers larger than those of TXS\,0506+056 in the IBL/HBL sample put together by \cite{Giommi_2021}. 
GB6\,J1542+6129 has $P_{\rm 1.4GHz} = 10^{26.4\pm0.7}$ W Hz$^{-1}$ and $L_{\gamma} = 
10^{47.2\pm0.9}$ erg s$^{-1}$.}. Since IBLs plus HBLs comprise $\sim 42$ per cent of the {\it Fermi}-4LAC (clean sample)
blazars with SED classification \citep{Ajello_2020,Lott_2020}, blazars of the type IceCube has associated with neutrinos constitute 
{\it at most} 1 per cent of the $\gamma$-ray selected population ($\lesssim 20$ sources), since we have
not included in our calculation their other peculiar properties.


\section*{Acknowledgments}
We thank Maria Petropoulou for her comments on the paper. This work is supported by the Deutsche
Forschungsgemeinschaft through grant SFB\,1258 ``Neutrinos and Dark Matter
in Astro- and Particle Physics''. 

\section*{Data Availability}
The flux-calibrated and de-reddened spectrum is available in the online database ZBLLAC ({\url{http://web.oapd.inaf.it/zbllac/}}).

\label{lastpage}

\bsp	

\end{document}